\documentclass[epj,nopacs,twocolumn]{svjour}

\usepackage[english]{babel}
\usepackage{amsmath}
\usepackage{graphicx} 
\usepackage{hyperref}
\usepackage{cite}
\usepackage{upgreek}

\begin{document}

\title{The emergence of local wrinkling or global buckling in thin freestanding bilayer films }

\author{John F. Niven \inst{1} \and Gurkaran Chowdhry \inst{1} \and James S. Sharp \inst{2} \and Kari Dalnoki-Veress \inst{1}$^{\text{,}}$ \inst{3}$^{\text{,}}$\thanks{\email{dalnoki@mcmaster.ca}}}
\institute{Department of Physics \& Astronomy, McMaster University, Hamilton, Ontario, L8S 4M1, Canada \and School of Physics and Astronomy, University of Nottingham, University Park, Nottingham, NG7 2RD, UK \and UMR CNRS Gulliver 7083, ESPCI Paris, PSL Research University, 10 rue Vauquelin, 75005 Paris, France}

\date{\today}

\abstract{Periodic wrinkling of a rigid capping layer on a deformable substrate  provides a useful method for templating surface topography for a variety of novel applications. Many experiments have studied wrinkle formation during the compression of a rigid film on a relatively soft pre-strained elastic substrate, and most have focused on the regime where the substrate thickness can be considered semi-infinite relative to that of the film. As the relative thickness of the substrate is decreased, the bending stiffness of the film  dominates, causing the bilayer to transition to either local wrinkling or a global buckling instability. In this work optical microscopy was used to study the critical parameters that determine the emergence of local wrinkling or global buckling of freestanding bilayer films consisting of a thin rigid polymer capping layer on a pre-strained elastomeric substrate. The thickness ratio of the film and substrate as well as the pre-strain were controlled and used to create a buckling phase diagram which describes the behaviour of the system as the ratio of the thickness of the substrate is decreased. A simple force balance model was developed to understand the thickness and strain dependences of the wrinkling and buckling modes, with excellent quantitative agreement being obtained with experiments using only independently measured material parameters.}

\maketitle

\section{Introduction}

Wrinkling and buckling of thin films has been thoroughly investigated for a variety of applications such as small-scale surface patterning \cite{JG2006, AC2008, YC2014}, biomedical devices \cite{YL2017}, and flexible electronics \cite{JR2010}. One way to achieve surface patterns is by capping a soft, stretchable substrate, such as an elastomer (elastic modulus $E_{\text{s}} \sim$ MPa) with a relatively thin rigid layer, such as a metallic or polymeric film ($E_{\text{f}} \sim$ GPa). If the rigid layer becomes sufficiently compressed it can buckle locally out of plane with a sinusoidal wrinkling pattern in order to accommodate its excess surface area relative to the compressed substrate. The out of plane deformation comes at the cost of bending the rigid capping layer in exchange for relaxing energy stored in compression. Compression can be achieved through either differential thermal expansion between the capping layer and the substrate material \cite{NB1998}, chemical swelling \cite{EC20061, EC20062, DB2011}, or by mechanically pre-straining the substrate prior to capping with the rigid film \cite{AV2000, CS2004, KE2005, PL2007}. The relationship between wrinkling wavelength and the system parameters has been so well established that it has been used as a metrology~\cite{ CS2004}.

Mechanically induced buckling, shown schematically in fig. \ref{fig:wrinkle_schem}, is most relevant for biomedical devices and flexible and wearable electronics applications, where the use of high temperatures or chemical swelling should be avoided \cite{YS2006, JR2010, MK2013}. In such systems it is critical to understand how mechanical instabilities and failure modes depend on the geometry of the  bilayer and the material properties of the individual layers. Most experiments to date have focused on macroscopic samples, where the thickness of the substrate, $H$, is taken to be infinite relative to the thickness of the capping film, $h$. Following release of the pre-strain, these systems can form either 1D or 2D wrinkling patterns \cite{PL2007}. At higher strains, localization features such as folds and creases \cite{LP2008, FB2013, YC2014, QW2015}, period doubling \cite{FB2010}, or delaminations can occur. In the latter case, voids form between the film and substrate \cite{HM2007, YE2012, AN2017}. In this semi-infinite regime, the wavelength of wrinkling, $\lambda$, can be calculated using a force balance between the bending of the capping film and the deformation of the substrate \cite{AV2000}. This force balance results in a wrinkle wavelength:
\begin{equation}
\lambda = 2\pi h \left(\frac{\bar{E}_{\text{f}}}{3\bar{E}_{\text{s}}}\right)^{1/3},
\label{lambda_inf}
\end{equation}
where $\bar{E}_i = E_i/(1-\nu_i^2)$ is the plane strain modulus of layer $i$, with $E_i$ the Young's modulus and $\nu_i$ the Poisson's ratio, where subscript $i$ = $f$ or $s$ refers to the capping film or substrate, respectively. This wavelength is linearly dependent on the thickness of the capping film, while being independent of the substrate thickness. This relationship has been proven exhaustively by experiments \cite{AV2000, JG2006} and allows wrinkling in the semi-infinite regime, $h<<H$, to be used as a method for measuring the moduli of thin rigid films \cite{CS2004,CS2005,CS2006,JC2011}. There is also a critical pre-strain, $\epsilon_{\text{c},\infty}$, required for wrinkling in the semi-infinite regime, which is dependent only on the ratio of the moduli of the materials:
\begin{equation}
\epsilon_{\text{c},\infty} = \frac{1}{4} \left(\frac{3\bar{E}_{\text{s}}}{\bar{E}_{\text{f}}}\right)^{2/3}.
\label{eq:ec_inf}
\end{equation}
Below this critical pre-strain the capping film remains flat upon compression. For typical elastomer/polymer material pairs this critical pre-strain is $\sim 0.5$~\% \cite{YE2012, YC2014}.

When considering applications such as flexible electronics or biomedical applications, technology continues to push for ever smaller structures and feature sizes, requiring bilayer films in which the thickness of the substrate can no longer be considered semi-infinite. In this finite-substrate regime it has been suggested both theoretically \cite{SW2008, YM20162, XM2017} and experimentally \cite{DH2009} that wrinkling wavelengths can deviate from those predicted by eq. \ref{lambda_inf}, and that this could affect device performance. Crucially, when the bending stiffness of the capping film becomes dominant, the entire bilayer film could undergo global buckling on a length scale similar to the total sample length, as shown schematically in fig. \ref{fig:wrinkle_schem}(b), similar to observations in refs. \cite{AT2011} and \cite{AC2007}. For a detailed theory for the characteristic wavelength associated with the global buckling, we refer the reader to the work by Takei and co-workers \cite{AT2011}.
\begin{figure}[b]
\centering
\includegraphics[width=0.5\textwidth,trim= 0cm 0cm 3cm 0cm]{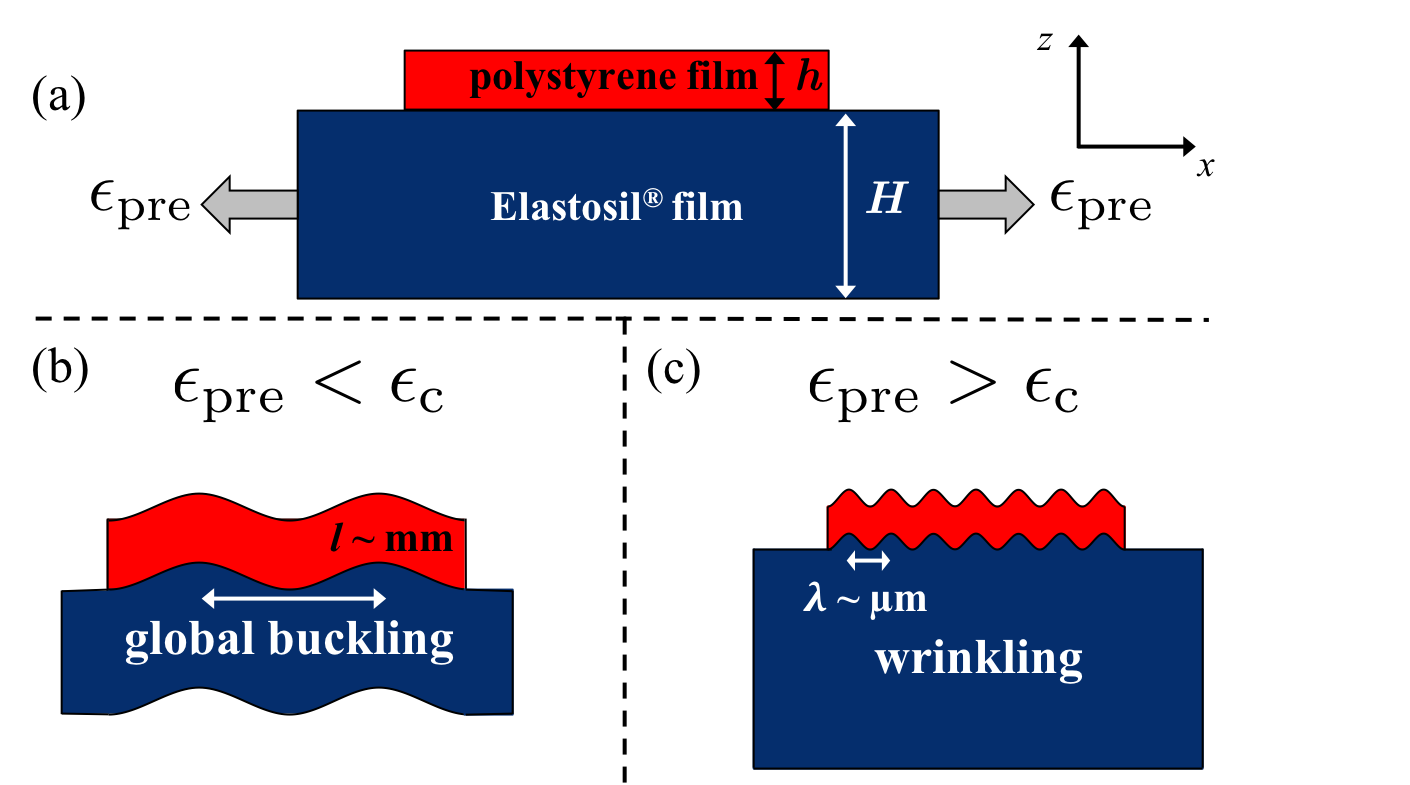}
\caption{(a) Side view schematic of an elastomeric film pre-strained by an amount $\epsilon_{\text{pre}}$, the direction of which is indicated by the grey arrows. The elastomer film is capped by a thin polystyrene film. (b) Below a critical pre-strain the bilayer will undergo buckling. (c) Above the critical pre-strain the rigid film will wrinkle.}
\label{fig:wrinkle_schem}
\end{figure} 
The critical conditions that separate local wrinkling and global buckling in bilayer films have been studied both theoretically and experimentally in the context of flexible electronics using macroscopic samples \cite{SW2008, YM2016, YM20162}. These works have developed theories to predict the type of instability observed with unclamped boundary conditions, and have predicted that the critical pre-strain required for wrinkling increases as the relative thickness of the substrate is decreased, although experimental results are limited. Previous work has not addressed what instability results, provided that the pre-strain is sufficient to destabilize the free-standing bilayer. If the bonding between the layers is strong then two possible scenarios emerge: global large wavelength buckling (fig.~\label{fig:wrinkle_schem} (b)) and local wrinkling (fig.~\label{fig:wrinkle_schem} (c)).

In this work we present an experiment to study the critical geometrical parameters which determine the morphology of a sufficiently pre-strained free-standing bilayer film which, upon relaxation, transitions into either local wrinkling or global buckling. The use of a freestanding geometry means that the entire bilayer sample is able to deform out of plane, allowing for global buckling to occur. By varying the thickness ratio of the capping film and substrate, $h/H$, and the substrate pre-strain, $\epsilon_{\text{pre}}$, the separation between the emergence of local wrinkling or global-buckling can be mapped out. In all our experiments we operate with pre-strains that are well above the critical pre-strain, $\epsilon_{\text{c},\infty}$. We also present a model, adapted from the semi-infinite theory, to produce a simple description of the separation between wrinkling and global buckling which is in good quantitative agreement with experiments.

\section{Experimental Methods}

\begin{figure}[t]
\centering
\includegraphics[width=0.49\textwidth,trim= 0cm 0cm 0cm 0cm]{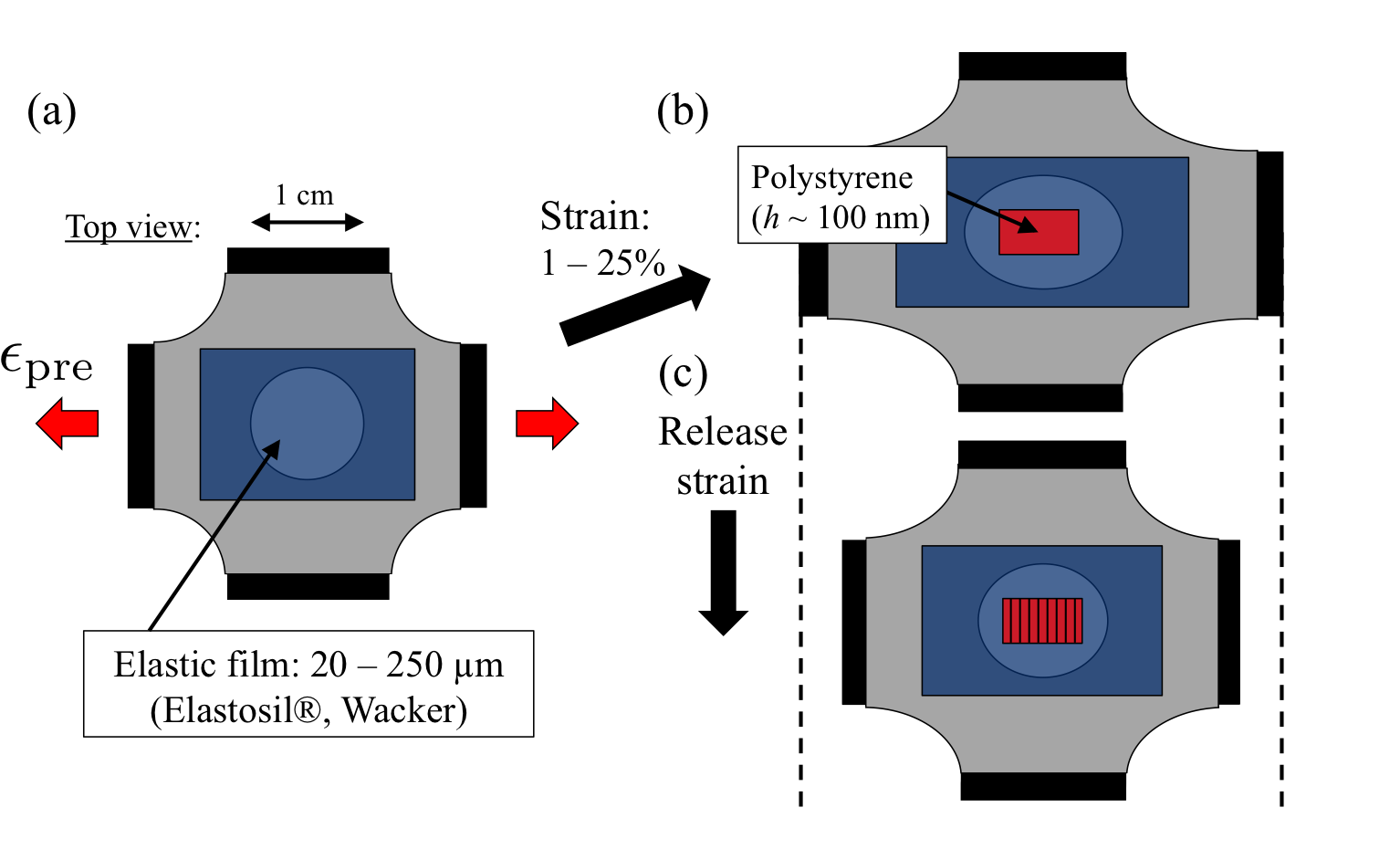}
\caption{(a) Top view schematic of the experimental straining setup showing the rounded plus sign 250 $\mu$m Elastosil film with a hole in the middle. The hole is covered with a second freestanding Elastosil film of varying thickness, which acts as the substrate. (b) The substrate is strained biaxially and capped with a thin PS film, forming a bilayer. (c) As strain is released and the bilayer is compressed, it can either wrinkle or buckle.}
\label{fig:exp_schem}
\end{figure}

Bilayer films were prepared on a biaxial straining apparatus shown schematically in fig. \ref{fig:exp_schem} and described previously \cite{BDP2018, RS2018}. The apparatus consisted of a 258 $\pm$ 2 $\mu$m thick Elastosil$^{\text{\textregistered}}$ sheet (Wacker Chemie AG), which is cut into a rounded ``plus" shape, with a 1 cm diameter hole in the middle. This shape was chosen to ensure nearly uniform biaxial strain at the centre of the hole where the bilayer sample is placed. We stress that this 258~$\mu$m sheet merely acts as a frame that is used to mount the bilayer sample of interest and apply a controlled  strain. Each arm of the ``plus'' shaped frame was clamped to a post and attached to crossed optical rails using translation stages.  The hole was covered with a second Elastosil film, which becomes the ``substrate" of the bilayer, with thicknesses of $H$ = \{20.9 $\pm$ 0.4, 51 $\pm$ 1, 104 $\pm$ 2, 213 $\pm$ 7, or 258 $\pm$ 2\} $\mu$m as measured in reference~\cite{JCODB2019}. Samples with intermediate substrate thickness were made by stacking Elastosil sheets with good adhesion between the films -- sufficient that the films remained in good contact when strained. The 0 \% strain value was calibrated before each experiment by adjusting each post until the point just before the Elastosil sheet begins to wrinkle. The substrate was then strained biaxially by moving two of the translation stages in opposite directions (the high-strain direction) while leaving the perpendicular direction fixed. The applied pre-strain was measured optically using $\epsilon_{\text{pre}} = (d_{\text{f}}-d_{\text{i}})/d_{\text{i}}$, where $d_{\text{i}}$ and $d_{\text{f}}$ are the initial and final distances between two defects in the film surface aligned parallel to the strain direction, respectively. The posts perpendicular to the applied strain were not adjusted, meaning that the strain in the perpendicular direction is zero. We note that while there is no strain in the perpendicular direction, there is a tensile stress induced through Poisson's ratio which is smaller than that induced in the high-strain direction. The Elastosil films have a modulus $E_{\text{s}}$ = 1.11 $\pm$ 0.06 MPa \cite{JCODB2019}, and making the reasonable assumption that the elastomer is incompressible then Poisson's ratio can be taken to be $\nu_{\text{s}} = 0.5$. 

Films of polystyrene (PS, with number averaged molecular weight $M_{\text{n}}$ = 185 kg mol$^{-1}$ and polydispersity index, PDI = 1.06, Polymer Source Canada, Poisson's ratio $\nu_{\text{f}} = 0.33$ \cite{polyhandbook} and modulus $E_{\text{f}}$ = 3.3 GPa \cite{polyhandbook}) with thickness values ranging from $h \approx$ 80 \--- 1900 nm were prepared by spin coating from dilute toluene solution onto freshly cleaved mica substrates and their thickness measured using ellipsometry (Accurion, EP3). We note that the modulus of PS at these thicknesses should remain at the bulk value \cite{CS2006}. All films were annealed in vacuum at 140~$^{\circ}$C for a minimum of several hours to relax the polymer chains and remove any residual solvent. The films were then cut into approximately 3 mm~$\times$~3 mm pieces using a scalpel and floated off the mica and onto the surface of an ultrapure water bath (18.2 M$\Upomega.$cm, Pall, Cascada, LS). A piece of the PS film was then floated back onto the mica substrate, and brought into contact with the Elastosil substrate. The strong adhesion between the PS and Elastosil means that the mica substrate can be removed, resulting in a free-standing PS/Elastosil bilayer. Any remaining water is gently wicked away at the edge of the sample. The pre-strain is then slowly released by moving the translation stages, and the bilayer film can be observed using optical microscopy as the capping film is compressed by the relaxing substrate. An important feature to note with this setup is that the bilayer film is freestanding in the region of the hole, so the entire film is free to deform out of plane of the compression.

\section{Results and Discussion}
\begin{figure}[t]
\centering
\includegraphics[width=0.5\textwidth,trim= 0cm 0cm 0cm 0cm]{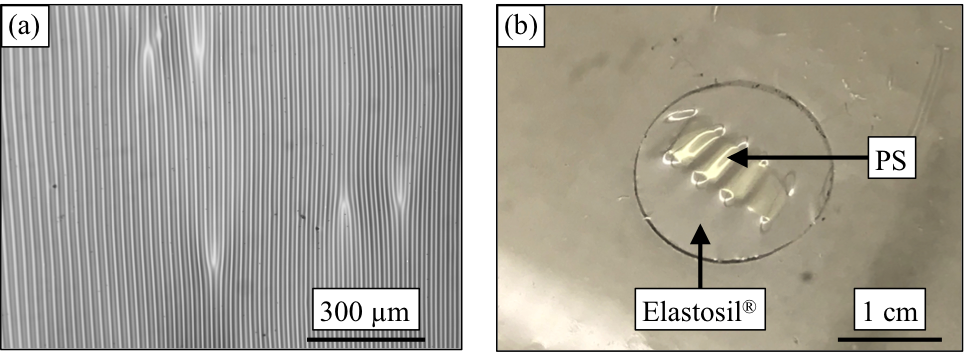}
\caption{(a) Optical microscopy image of wrinkles in a PS/Elastosil bilayer. The wrinkling wavelength varies between 1 - 100 $\mu$m depending on the capping film thickness. (b) Optical image of a buckled bilayer film with a length scale $\sim$mm.}
\label{fig:wrinklepics}
\end{figure}

\begin{figure}[b]
\centering
\includegraphics[width=0.5\textwidth,trim= 0cm 0cm 0cm 0cm]{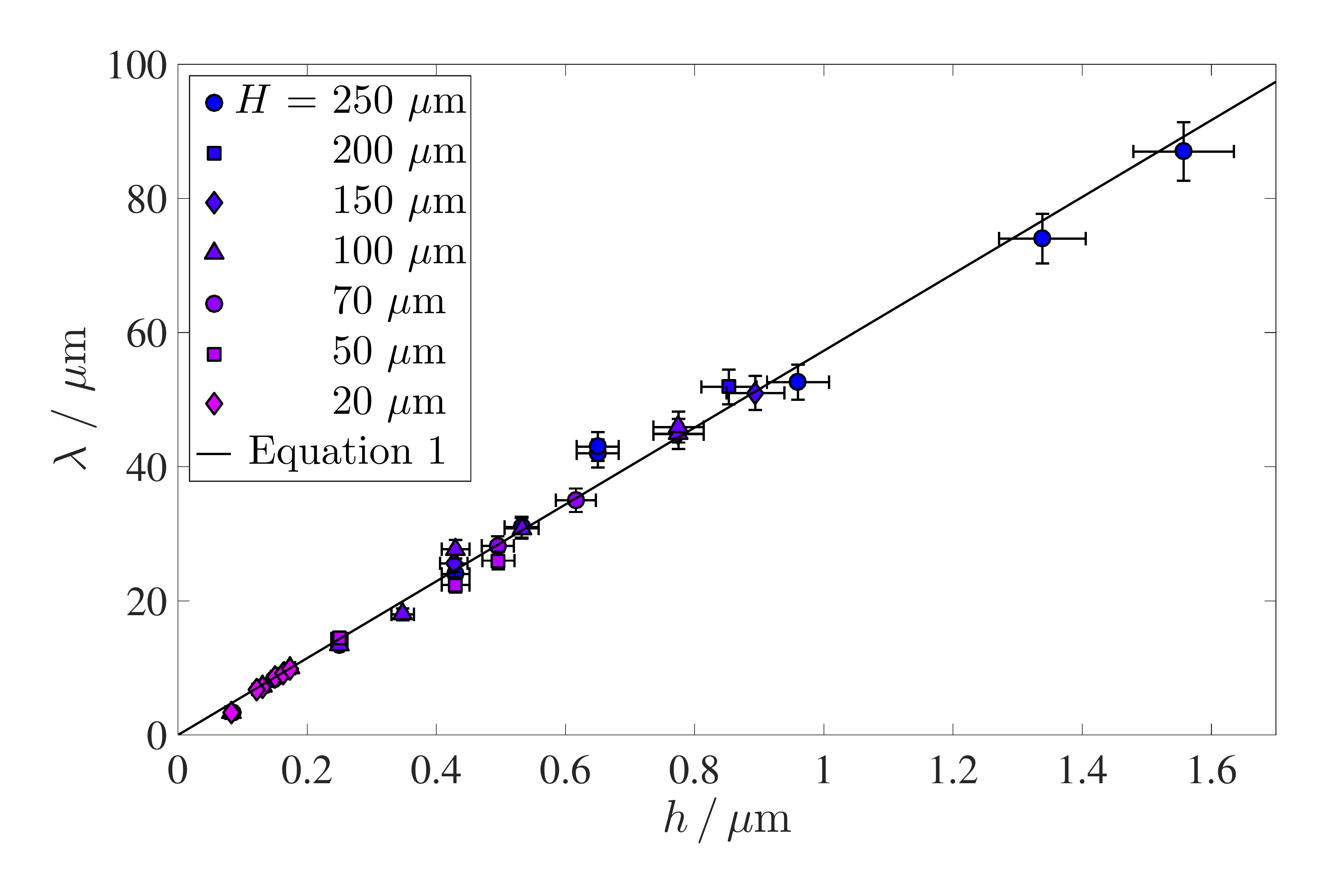}
\caption{Wrinkling wavelength, $\lambda$, as a function of the capping film thickness, $h$, for a range of substrate thicknesses, $H$, for pre-strains up to 25\% and 3.4$\times10^{-4} < h/H < 8.7\times10^{-3}$. The solid line corresponds to the semi-infinite theory, eq. \ref{lambda_inf}, with no fitting parameters.}
\label{fig:wavelength}
\end{figure}

Wrinkling was observed in this system with wavelengths ranging between 1 $\mu$m and 100 $\mu$m (as shown in fig. \ref{fig:wrinklepics}(a)), as measured using optical microscopy. Figure \ref{fig:wavelength} shows the measured wavelengths as a function of the rigid capping film thickness, $h$, for substrate thicknesses ranging from $H$ = 20 $\mu$m to 250~$\mu$m for strains up to 25\%, as well as a line corresponding to the semi-infinite model, eq. \ref{lambda_inf}. We can see that, even with different substrate film thicknesses, the data is in excellent agreement with the semi-infinite model with no fitting parameters, only independently measured moduli and Poisson's ratios of the capping film and elastic substrate. This result indicates that, provided that wrinkling is observed, for these materials at pre-strains between 1 -- 25\%, the semi-infinite model is valid for 3.4$\times10^{-4} < h/H < 8.7\times10^{-3}$.

For a given pair of materials, the semi-infinite wrinkling regime exists for low thickness ratios, $h/H \approx 1\times10^{-3}$, and above the critical strain. At a fixed pre-stain, as $h/H$ is increased there is a critical value at which the observed instability upon release of the strain changes from local wrinkling, as seen in fig. \ref{fig:wrinklepics}(a), to global buckling instability (shown in fig. \ref{fig:wrinklepics}(b)), with a length scale on the order of the total sample length.

In order to obtain a quantitative understanding of the critical conditions that separate local wrinkling and global buckling in this system, we develop a simple model beginning with a force balance between the bending of the rigid capping film and the deformation of the substrate \cite{LandL}. Similar derivations have been shown in the literature previously, but here we present the derivation alongside that for buckling for completeness \cite{MB1937, HA1969, AV2000}. The force balance in the bilayer can be written in the form:

\begin{equation}
\bar{E}_{\text{f}}I_{\text{f}}z'''' + \bar{E}_{\text{s}} \frac{wq}{2}z + F_{\text{w}} z'' = 0,
\label{eq:eqn1}
\end{equation}
where the first term considers the bending of the PS film, the second describes the forces in the substrate and the final term deals with the in-plane forces in the PS film that are generated by the release of the pre-strain in the substrate.  

\begin{figure}[t]
\centering
\includegraphics[width=0.5\textwidth,trim= 0cm 0cm 0cm 0cm]{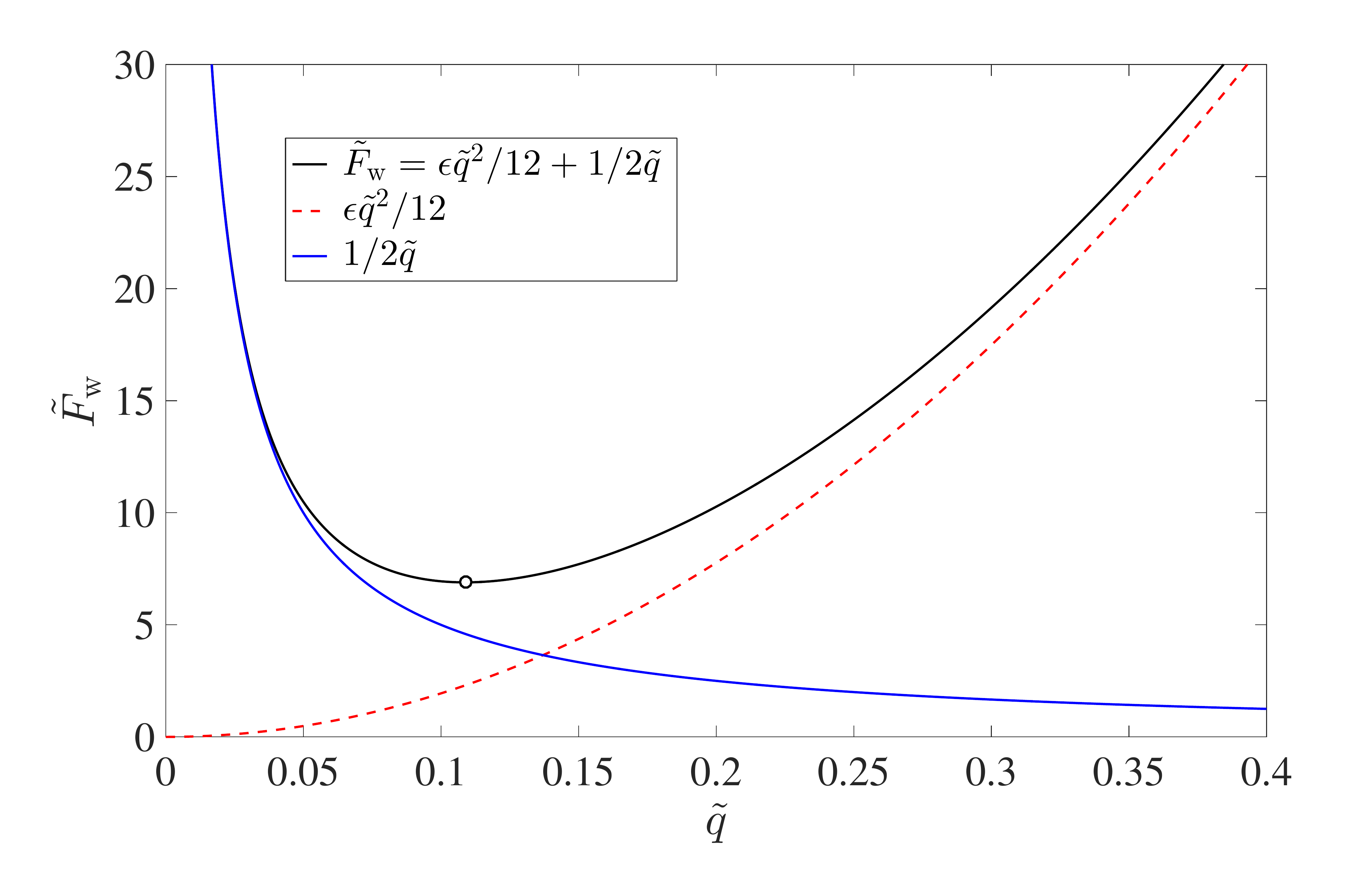}
\caption{Plot of eq. \ref{eq:wrinkle} showing the non-dimensional force, $\tilde{F}_{\text{w}}$, required for wrinkling as a function of the non-dimensional wavenumber $\tilde{q}=qh$, shown in black. The dashed red line and the blue line represent the first and second term on the RHS of eq. \ref{eq:wrinkle}. The black circle shows the minimum  force required for wrinkling, $\tilde{F}_{\text{w, c}}$.}
\label{fig:force}
\end{figure}

The deformation of the top surface of the bilayer is given by:
\begin{equation}
z(x) = A\sin{qx},
\label{eq:z}
\end{equation}
where $q = 2\pi/\lambda$ is the wavenumber. Primes in eq. \ref{eq:eqn1} denote derivatives with respect to the in-plane direction $x$, $I_{\text{f}} = wh^3/12$ is the inertial moment of the capping film of thickness $h$ and width $w$, and $F_{\text{w}}$ is the force applied to the capping film which causes wrinkling. Differentiating eq. \ref{eq:z} and substituting into eq. \ref{eq:eqn1} gives:
\begin{equation}
\frac{F_{\text{w}}}{w} = \frac{\bar{E}_{\text{f}} h^3 q^2}{12} + \frac{\bar{E}_{\text{s}}}{2q}.
\end{equation}
This equation can be simplified by non-dimensionalizing as follows: $\tilde{F}_{\text{w}} \equiv \frac{{F}_{\text{w}}}{wh\bar{E}_{\text{s}}}$, $\tilde{\lambda} \equiv \frac{\lambda}{h}$, $\tilde{q} \equiv qh$, $\tilde{H} \equiv \frac{H}{h}$, $\gamma \equiv \frac{\bar{E}_{\text{f}}}{\bar{E}_{\text{s}}}$, which results in the following condition for wrinkling:
\begin{equation}
\tilde{F}_{\text{w}} = \frac{\gamma\tilde{q}^2}{12} + \frac{1}{2\tilde{q}}.
\label{eq:wrinkle}
\end{equation}
Equation \ref{eq:wrinkle} can be minimized to get the relevant critical values for wrinkling:
\begin{eqnarray}
\tilde{q}_{\text{c}} &=& \left( \frac{3}{\gamma}\right)^{1/3}, \\
\tilde{\lambda}_{\text{c}} &=& 2\pi \left( \frac{\gamma}{3}\right)^{1/3}, \\
\mathrm{ and \ }\tilde{F}_{\text{w,c}} &=& \frac{3}{4} \left( \frac{\gamma}{3}\right)^{1/3}.
\end{eqnarray}
Figure \ref{fig:force} shows a plot of the critical force required for wrinkling, eq. \ref{eq:wrinkle}, including the two terms, as a function of $\tilde{q}$. The minimum critical force for wrinkling is shown as a circle.

The criterion for wrinkling is simply that the pre-strain in the substrate is such that the force applied to the capping film, $F_{\text{pre}}$, is enough to overcome the critical force $F_{\text{w,c}}$. If there is not sufficient force on the capping film then the bilayer can only buckle or remain flat. As the strain is relaxed, the force due to the pre-strain increases slowly and reaches a maximum when the substrate pre-strain in the region outside of the capping film is a minimum. Thus the wrinkling criterion now becomes $F_{\text{pre}} = F_{\text{w,c}}$. The maximum force resulting from the initial pre-strain (i.e. when there is full relaxation of the uncapped region of the substrate) is given by $F_{\text{pre}} = \epsilon_{\text{pre}} w H \bar{E}_{\text{s}}$, since $\epsilon = \frac{\sigma_c}{\bar{E}_f} = \frac{F_c}{hw\bar{E}_f}$. Therefore: $F_{\text{pre}} = \epsilon_{\text{pre}} w H \bar{E}_{\text{s}} = F_{\text{w,c}} = \tilde{F}_{\text{w,c}} \bar{E}_{\text{s}}w h$, which results in:

\begin{equation}
h = \frac{4}{3}\cdot \left(\frac{3\bar{E}_{\text{s}}}{\bar{E}_{\text{f}}}\right)^{1/3}\epsilon_{\text{pre}} H.
\label{eq:transition}
\end{equation}

Instead of taking the maximum force as resulting from the applied pre-strain alone, we recognize that even with zero applied pre-strain the act of transferring the PS capping film onto the freestanding substrate membrane causes it to deform, which induces an additional pre-strain that remains after the bilayer is formed. To test this, a bilayer was made with ``0 \%" applied strain using the same calibration technique as all other experiments. When compressed, the sample had regions of clear wrinkling throughout, which is only possible if there was indeed a small pre-strain induced from the sample preparation process. Additionally, there is some uncertainty in the true ``0 \%" pre-strain value because the boundary between a tensile strain and compressive strain was determined by the appearance of small wrinkles in the substrate by eye.  If we assume that this induced deformation is consistent between experiments, the total strain in the substrate is then: $\epsilon_{\text{total}} = \epsilon_{\text{pre}} + \epsilon_0$, where $\epsilon_0$ is a small additional pre-strain resulting from the transfer of the PS film. It then follows that $F_{\text{pre}} = (\epsilon_{\text{pre}} + \epsilon_0) w H E_{\text{s}}$. Balancing this with the force required for wrinkling gives the following criterion for wrinkling:
\begin{equation}
h = \frac{4}{3}\cdot \left(\frac{3\bar{E}_{\text{s}}}{\bar{E}_{\text{f}}}\right)^{1/3}\left(\epsilon_{\text{pre}} + \epsilon_0\right) H.
\label{eq:transition2}
\end{equation}

\begin{figure}[t]
\centering
\includegraphics[width=0.5\textwidth,trim= 0cm 0cm 0cm 0cm]{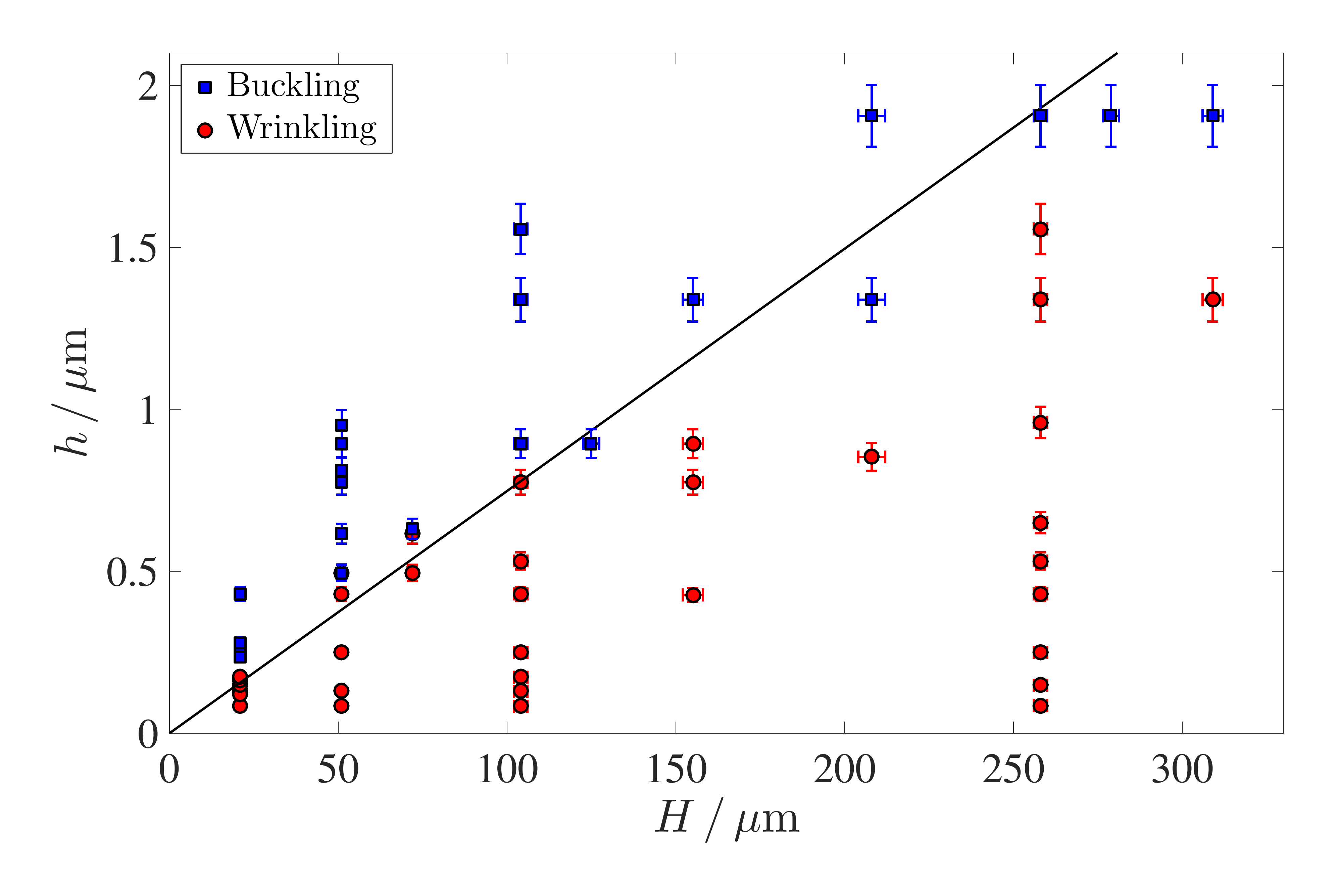}
\caption{Phase diagram of wrinkling (red circles) and buckling (blue squares) for various substrate and cap thickness at a fixed value of $\epsilon_{\text{pre}}$ = 3 \% (total pre-strain of $\epsilon_{\text{pre}} + \epsilon_0$ = 5.3 \%). The solid line corresponds to the theoretical transition between wrinkling and buckling, eq. \ref{eq:transition2}, with no fitting parameters.}
\label{fig:phase1}
\end{figure}

Experiments were carried out for various substrate and capping film thickness and at various pre-strain values which always exceeded the critical-pre-strain, $\epsilon_{\text{c},\infty}$. For samples where the pre-strain is below $\epsilon_{\text{c},\infty}$, the sample simple remains as a flat slab. The experimental strain offset was measured to be $\epsilon_0$ = 2.3 \%. Figure \ref{fig:phase1} shows the phase diagram of wrinkling and buckling at a fixed applied pre-strain value of 3 \% (total pre-strain of 5.3 \%) for various capping film and substrate thicknesses. Wrinkles dominate at low values of $h/H$ (the semi-infinite regime), and buckles become dominant for higher values of $h/H$. There is a clear linear transition between wrinkling and buckling, with the straight line corresponding to eq. \ref{eq:transition2}. This plot shows excellent agreement between the data and theory given that there are no fitting parameters to the value of the slope, only material properties.

Equation \ref{eq:transition2} can be rewritten instead as the critical pre-strain required for wrinkling:
\begin{equation}
\epsilon_{\text{pre}} + \epsilon_0 = \frac{3}{4}\cdot \left(\frac{\bar{E}_{\text{f}}}{3\bar{E}_{\text{s}}}\right)^{1/3} \cdot \frac{h}{H},
\label{eq:ec_finite}
\end{equation}
which deviates from the critical value in the semi-infinite regime, eq. \ref{eq:ec_inf}, in that the finite regime value depends inversely on the substrate thickness. The resemblance of eq.~\ref{eq:ec_finite} to eq. \ref{eq:ec_inf} is fortuitous and does not agree in the limit of $H \rightarrow \infty$ because they describe two different mechanisms. 

\begin{figure}[b]
\centering
\includegraphics[width=0.5\textwidth,trim= 0cm 0cm 0cm 0cm]{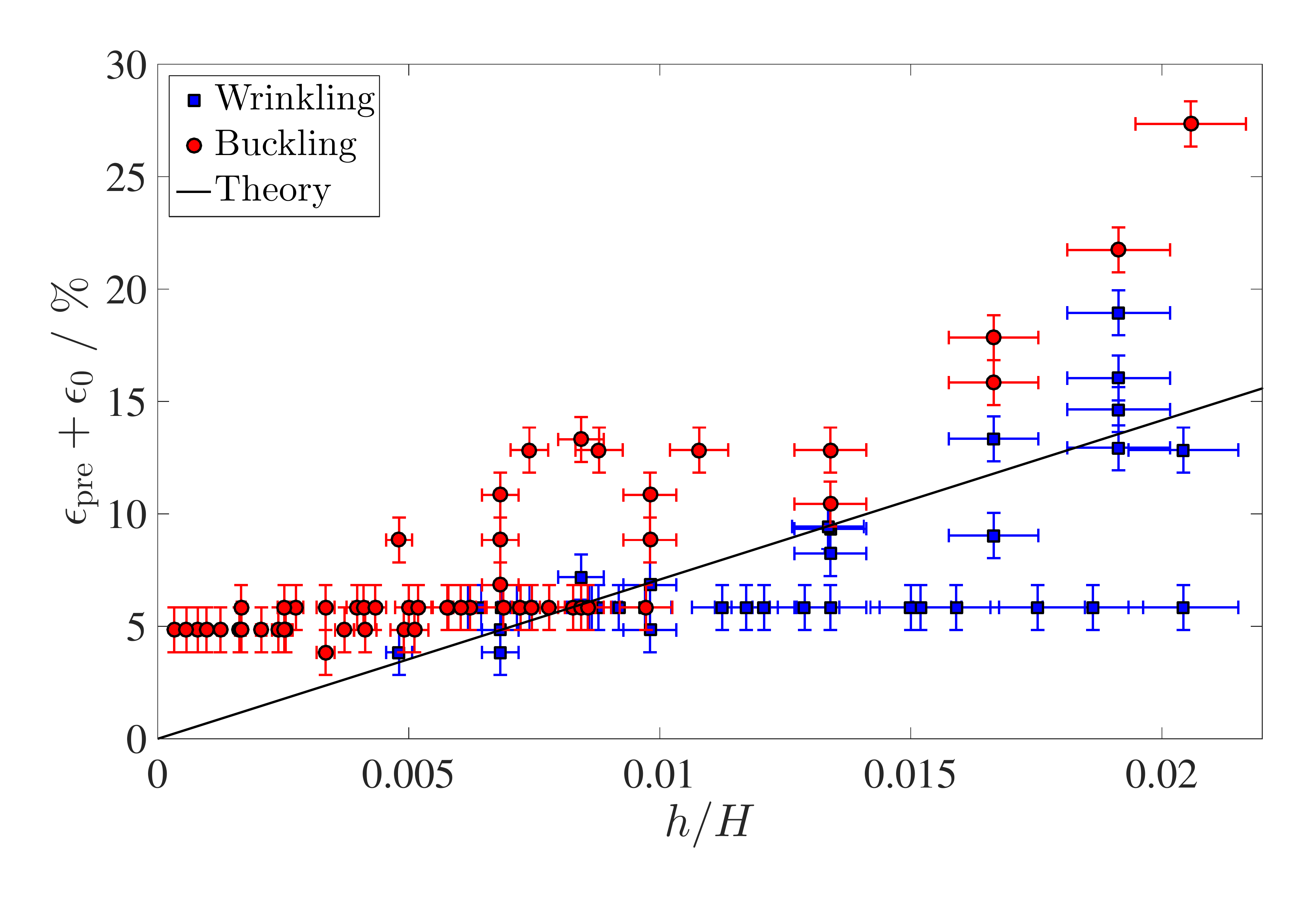}
\caption{Phase diagram of wrinkling (red circles) and buckling (blue squares) for various film thickness ratios, $h/H$, as a function of the full applied pre-strain $\epsilon_{\text{pre}} + \epsilon_0$. The solid line corresponds to the theoretical transition between wrinkling and buckling, eq. \ref{eq:ec_finite}, with no fitting parameters.}
\label{fig:phase2}
\end{figure}

Figure \ref{fig:phase2} shows the wrinkling and buckling phase diagram now with total pre-strain on the vertical axis and film thickness ratio $h/H$ on the horizontal axis. It is clear from this plot that for larger values of $h/H$ the system can be made to transition from buckling to wrinkling by increasing the pre-strain in the substrate, which increases the compressive force applied to the capping films. The solid line corresponds to eq. \ref{eq:ec_finite} and contains no fitting parameters. As this equation describes the transition from wrinkling to buckling in the bilayer samples, all data points below the line should correspond to samples that have buckled and those above the line to wrinkled samples. As the figure shows, there is excellent agreement between the theory and the experimental data for a wide range of thickness ratios and pre-strain values. There is significant uncertainty at low strains near the transition because samples would often show partial wrinkling, where only some of the area had wrinkled, which made distinguishing between the two regimes difficult. At pre-strains above 20 \% delaminations were found to become dominant as the buckling energy of the capping film overcame the adhesion energy between the films. This regime was not the focus of this study.

\section{Observed Delamination Morphologies}
While the delamination regime is beyond the scope of this manuscript, it is worth noting some of the morphologies that were observed, even if the results presented are merely observational.  It was found that interesting morphologies could be observed for high pre-strains in the semi-infinite regime, which included creases, folds, period doubling, and delamination. Novel post-delamination buckling morphologies have also been found under certain conditions, including ``varicose" or ``bubble" blisters \cite{MG2002, BA2002}, and ``telephone cord" blisters \cite{AB1999, MM2004, YN2017}. To study the effect of high strain in the finite substrate regime, bilayers with thickness ratios of order 1 were made using a different elastomeric film, Elastollan TPU 1185A (BASF). Elastollan films were prepared by spin coating from dilute cyclohexanone solution with thicknesses of $\sim$1 $\mu$m. An Elastollan film was transferred onto the straining setup and strained biaxially by 15 \%. The film was then capped with a 500 nm thick PS film as before. Upon compression, large scale global buckling of the entire bilayer was observed first (fig. \ref{fig:wrinkle_schem}a). On further compression the PS film delaminated from the substrate, with delaminations forming across the entire width of the sample perpendicular to the compression axis. The delaminations were observed both optically and using atomic force microscopy (AFM) to scan the PS/air and Elastollan/air interfaces. Prior to AFM imaging, the sample was transferred to a thin stainless steel washer. Elastollan is adhesive enough that the film remained in good contact with the washer without losing strain, and the washer is thin enough that the sample can simply be flipped in order to image both interfaces. A scan of the two interfaces shows that the delamination ridge is only present at the PS/air interface, meaning there is a void between the two films.

\begin{figure}[t]
\centering
\includegraphics[width=0.4\textwidth,trim= 0cm 0cm 0cm 0cm]{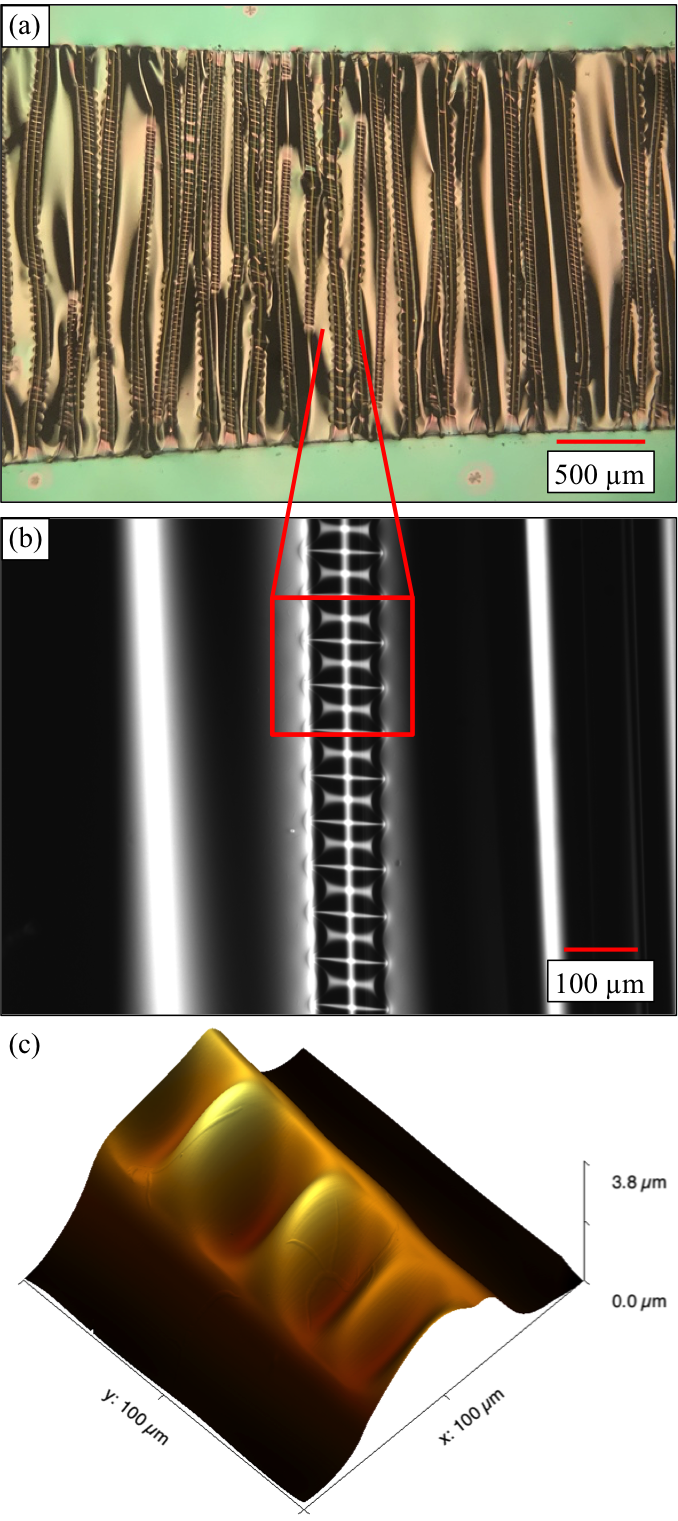}
\caption{(a) Optical microscopy image showing ``zipper" delaminations in a bilayer consisting of a 1 $\mu$m Elastollan film capped with a 500 nm PS film. (b) Zoomed in optical microscopy image in reflection mode showing the periodic structure of a zipper delamination. (c) AFM image of the Elastollan/air interface of a ``zipper" delamination showing the periodic structure}
\label{fig:zipper1}
\end{figure}

Due to the Poisson's ratio of the pre-strained elastomer there is also an initial tension in the elastic film orthogonal to the pre-strain direction, which cannot be relaxed while the film and substrate remain in good contact \cite{RS2018}. However, as the delaminations grow, contact is lost between the two films, and the substrate is now free to relax its excess length along the delaminated regime while remaining in contact along the edges. This geometry results in an instability with a periodic structure along the length of the delamination, seen optically in fig. \ref{fig:zipper1}(a) and (b). The structure was studied in more detail using AFM (fig. \ref{fig:zipper1}(c)) and showed that the periodic pattern is caused by wrinkling in the elastomer layer, while the PS film remains in its one-dimensional delaminated structure without buckling significantly in the perpendicular direction. The observed pattern is reminiscent of the ``bubble" delaminations seen previously \cite{MG2002, BA2002}, and provides a novel technique for templating freestanding films.

\section{Conclusions}
In conclusion, we have observed the transition between wrinkling and buckling in freestanding rigid/elastic bilayer films for which the substrate thickness cannot be taken as semi-infinite. We have shown that the critical pre-strain for wrinkling depends on film/substrate thickness ratio, $h/H$. A simple force balance model was used to predict the critical criteria required for wrinkling, which matches well with the experimental data using only independently measured material parameters. This model deviates from semi-infinite theory in that it as a dependence on the substrate thickness. These results provide experimental insights into design considerations for flexible electronics and other applications with thin elastic substrates.

\begin{acknowledgement}
Financial support was provided by the National Sciences and Engineering Research Council (NSERC). The authors thank Wacker Chemie AG for donating the Elastosil$^{\text{\textregistered}}$ material. JSS would like to acknowledge the support of the University of Nottingham?s International Collaboration Fund.
\end{acknowledgement}

\section*{Author contribution statement}
JN and KDV designed the research project, JN and GC performed all experiments and analyzed the data, JN, JS and KDV developed the theoretical model, JN wrote the first draft of the manuscript, and all authors edited the manuscript to generate a final version and contributed to the discussion throughout the entire process of the research.


\begin{thebibliography}{44}

\bibitem{JG2006}
J.~Genzer, J.~Groenewold, Soft Matter \textbf{2}, 310 (2006)

\bibitem{AC2008}
A.~Chiche, C.M. Stafford, J.T. Cabral, Soft Matter \textbf{4}, 2360 (2008)

\bibitem{YC2014}
Y.C. Chen, A.J. Crosby, Adv. Mater. \textbf{26}, 5626 (2014)

\bibitem{YL2017}
Y.~Liu, M.~Pharr, G.A. Salvatore, ACS Nano \textbf{11}, 9614 (2017)

\bibitem{JR2010}
J.A. Rogers, T.~Someya, Y.~Huang, Science \textbf{327}, 1603 (2010)

\bibitem{NB1998}
N.~Bowden, S.~Brittain, A.G. Evans, J.W. Hutchinson, G.M. Whitesides, Nature
  \textbf{393}, 146 (1998)

\bibitem{EC20061}
E.P. Chan, A.J. Crosby, Soft Matter \textbf{2}, 324 (2006)

\bibitem{EC20062}
E.P. Chan, A.J. Crosby, Adv. Mater. \textbf{18}, 3238 (2006)

\bibitem{DB2011}
D.~Breid, A.J. Crosby, Soft Matter \textbf{7}, 4490 (2011)

\bibitem{AV2000}
A.L. Volynskii, S.~Bazhenov, O.V. Lebedeva, N.F. Bakeev, J. Mater. Sci.
  \textbf{35}, 547 (2000)

\bibitem{CS2004}
C.M. Stafford, C.~Harrison, K.L. Beers, A.~Karim, E.J. Amis, M.R.
  VanLandingham, H.C. Kim, W.~Volksen, R.D. Miller, E.E. Simonyi, Nat. Mater.
  \textbf{3}, 545 (2004)

\bibitem{KE2005}
K.~Efimenko, M.~Rackaitis, E.~Manias, A.~Vaziri, L.~Mahadevan, J.~Genzer, Nat.
  Mater. \textbf{4}, 293 (2005)

\bibitem{PL2007}
P.C. Lin, S.~Yang, Appl. Phys. Lett. \textbf{90}, 241903 (2007)

\bibitem{YS2006}
Y.~Sun, V.~Kumar, I.~Adesida, J.~Rogers, Advanced Materials \textbf{18}, 2857
  (2006)

\bibitem{MK2013}
M.~Kaltenbrunner, T.~Sekitani, J.~Reeder, T.~Yokota, K.~Kuribara, T.~Tokuhara,
  M.~Drack, R.~Schw{\"o}diauer, I.~Graz, S.~Bauer-Gogonea et~al., Nature
  \textbf{499}, 458 EP  (2013)

\bibitem{LP2008}
L.~Pocivavsek, R.~Dellsy, S.~Johnson, B.~Lin, K.Y.C. Lee, E.~Cerda, Science
  \textbf{320}, 912 (2008)

\bibitem{FB2013}
F.~Brau, P.~Damman, H.~Diamant, T.A. Witten, Soft Matter \textbf{9}, 8177
  (2013)

\bibitem{QW2015}
Q.~Wang, X.~Zhao, Sci. Rep. \textbf{5}, 8887 (2015)

\bibitem{FB2010}
F.~Brau, H.~Vandeparre, A.~Sabbah, C.~Poulard, A.~Boudaoud, P.~Damman, Nat.
  Phys. \textbf{7}, 56 (2010)

\bibitem{HM2007}
H.~Mei, R.~Huang, J.Y. Chung, C.M. Stafford, H.H. Yu, Appl. Phys. Lett.
  \textbf{90}, 151902 (2007)

\bibitem{YE2012}
Y.~Ebata, A.B. Croll, A.J. Crosby, Soft Matter \textbf{8}, 9086 (2012)

\bibitem{AN2017}
A.J. Nolte, J.~Young~Chung, C.S. Davis, C.M. Stafford, Soft Matter \textbf{13},
  7930 (2017)

\bibitem{CS2005}
C.M. Stafford, S.~Guo, C.~Harrison, M.Y.M. Chiang, Rev. Sci. Instrum.
  \textbf{76}, 062207 (2005)

\bibitem{CS2006}
C.M. Stafford, B.D. Vogt, C.~Harrison, D.~Julthongpiput, R.~Hunag,
  Macromolecules \textbf{38}, 5095 (2006)

\bibitem{JC2011}
J.Y. Chung, A.J. Nolte, C.M. Stafford, Adv. Mater. \textbf{23}, 349 (2011)

\bibitem{SW2008}
S.~Wang, J.~Song, D.H. Kim, Y.~Huang, J.A. Rogers, Appl. Phys. Lett.
  \textbf{93}, 023126 (2008)

\bibitem{YM20162}
Y.~Ma, Y.~Xue, K.I. Jang, X.~Feng, J.A. Rogers, Y.~Huang, Proc. Royal Soc. A
  \textbf{472}, 20160339 (2016)

\bibitem{XM2017}
X.~Meng, G.~Liu, Z.~Wang, S.~Wang, Appl. Math. Mech. \textbf{38}, 469 (2017)

\bibitem{DH2009}
D.C. Hyun, U.~Jeong, J. Appl. Polym. Sci. \textbf{112}, 2683 (2009)

\bibitem{AT2011}
A.~Takei, F.~Brau, B.~Roman, J.~Bico, Europhys. Lett. \textbf{96}, 64001 (2011)

\bibitem{AC2007}
A.~Concha, J.W. McIver, P.~Mellado, D.~Clarke, O.~Tchernyshyov, R.L. Leheny,
  Phys. Rev. E \textbf{75}, 016609 (2007)

\bibitem{YM2016}
Y.~Ma, K.I. Jang, L.~Wang, H.N. Jung, J.W. Kwak, Y.~Xue, H.~Chen, Y.~Yang,
  D.~Shi, X.~Feng et~al., Adv. Funct. Mater. \textbf{26}, 5345 (2016)

\bibitem{BDP2018}
B.~Davis-Purcell, P.~Soulard, T.~Salez, E.~Rapha{\"e}l, K.~Dalnoki-Veress, Eur.
  Phys. J. E \textbf{41}, 36 (2018)

\bibitem{RS2018}
R.D. Schulman, J.F. Niven, M.A. Hack, C.~DiMaria, K.~Dalnoki-Veress, Soft
  Matter \textbf{14}, 3557 (2018)

\bibitem{JCODB2019}
J.C. Ono-dit Biot, M.~Trejo, E.~Loukiantcheko, M.~Lauch, E.~Rapha\"el,
  K.~Dalnoki-Veress, T.~Salez, Phys. Rev. Fluids \textbf{4}, 014808 (2019)

\bibitem{polyhandbook}
J.~Brandup, E.~Immergut, G.~EA, eds., \emph{Polymer Handbook}, Vol.~49,
  4th~edn. (Wiley and Sons, New York, 1999)

\bibitem{LandL}
L.D. Landau, E.M. Liftshitz, \emph{Theory of Elasticity}, 3rd~edn.
  (Butterworth-Heinemann, New York, USA, 1986)

\bibitem{MB1937}
M.~Biot, Journal of Applied Mechanics p.~A1 (1937)

\bibitem{HA1969}
H.~Allen, \emph{Analysis and Design of Structural Sandwich Panels}, 1st~edn.
  (Pergamon, 1969)

\bibitem{MG2002}
M.~George, C.~Coupeau, J.~Colin, F.~Cleymand, J.~Grilh{\'e}, Philos. Mag. A
  \textbf{82}, 633 (2002)

\bibitem{BA2002}
B.~Audoly, B.~Roman, A.~Pocheau, Eur. Phys. J. B \textbf{27}, 7 (2002)

\bibitem{AB1999}
B.~Audoly, Phys. Rev. Lett. \textbf{83}, 4124 (1999)

\bibitem{MM2004}
M.W. Moon, K.R. Lee, K.~Oh, J.~Hutchinson, Acta Mater. \textbf{52}, 3151
  (2004)

\bibitem{YN2017}
Y.~Ni, S.~Yu, H.~Jiang, L.~He, Nat. Commun. \textbf{8}, 14138 (2017)

\end{thebibliography}
\end{document}